\documentclass[a4paper,fleqn,usenatbib]{mnras}

\usepackage{newtxtext,newtxmath}

\usepackage[T1]{fontenc}
\usepackage{ae,aecompl}

\usepackage{array}
\usepackage{graphicx}	
\usepackage{amsmath}	
\usepackage{amssymb}	

\title[Spectral study of XTE J1710-281]{Broad-band spectral analysis of LMXB XTE J1710-281 with \emph{Suzaku}}
\author[P. Sharma et al.]{
Prince Sharma,$^{1}$\thanks{E-mail: princerajsharma31@gmail.com}
Rahul Sharma,$^{1}$
Chetana Jain$^{2}$
and Anjan Dutta$^{1}$
\\
$^{1}$Department of Physics and Astrophysics, University of Delhi, Delhi 110007, India\\
$^{2}$Hansraj College, University of Delhi, Delhi 110007, India\\
}

\date{Accepted XXX. Received YYY; in original form ZZZ}

\pubyear{2020}

\begin{document}
\label{firstpage}
\pagerange{\pageref{firstpage}--\pageref{lastpage}}
\maketitle

\begin{abstract}
This work presents the broad-band time-averaged spectral analysis of neutron star low-mass X-ray binary, XTE J1710-281 by using the \emph{Suzaku} archival data. The source was in a hard or an intermediate spectral state during this observation. This is the first time that a detailed spectral analysis of the persistent emission spectra of XTE J1710-281 has been done up to 30 keV with improved constraints on its spectral parameters. By simultaneously fitting the XIS (0.6--9.0 keV) and the HXD-PIN (15.0--30.0 keV) data, we have modelled the persistent spectrum of the source with models comprising a soft component from accretion disc and/or neutron star surface/boundary layer and a hard Comptonizing component. The 0.6--30 keV continuum with neutral absorber can be described by a multi-colour disc blackbody with an inner disc temperature of $kT_{\rm disc} = 0.28$ keV, which is significantly Comptonized by the hot electron cloud with electron temperature of $kT_{\rm e} \approx 5$ keV and described by photon index $\Gamma = 1.86$. A more complex three-component model comprising a multi-colour disc blackbody $\approx 0.30$ keV, single temperature blackbody $\approx 0.65$ keV and Comptonization from the disc, partially absorbed (about 38 per cent) by an ionized absorber (log($\xi$) $\approx$ 4) describes the broad-band spectrum equally well.

\end{abstract}




\section{Introduction}

Neutron Star (NS) Low-Mass X-ray Binaries (LMXBs) are binary star systems which consist of a weakly magnetized NS accreting from a low-mass ($\lesssim$ 1 M$_{\sun} $) companion star \citep{Lewin1980}. The accretion onto the NS from the companion star occurs through Roche-lobe overflow when the companion star evolves and expels material. Accretion takes place through the formation of an accretion disc around the NS with inner region of the disc heating up to $\sim 10^{7}$ K to emit X-ray radiation \citep{Shakura1973}. 

\citet{Mitsuda1984}, \citet{White1985}, and \citet{Mitsuda1989} evinced the bimodality in the X-ray spectral states of LMXBs, namely the high/soft state and the low/hard state. The canonical model, describing the X-ray spectra of LMXBs (e.g., Sco X-1, 4U 1608-522, GX 5-1 and GX 349+2) in soft state, consists of a multi-colour blackbody component attributed to the emission from accretion disc near the NS and a blackbody component to account for the emission from NS surface/boundary layer ($kT \sim$ 0.5--2 keV) with a weak ($\sim$ 3--5 keV) Comptonization \citep{Mitsuda1984,Mitsuda1989,Iaria2019}. This emission model has been strengthened by many studies \citep{Zhang2014,Zhang2016,Gambino2019}. 

During the hard spectral state, the emission spectrum is rather hard extending up to tens of keV due to the Comptonization of soft photons in hot optically thin corona \citep{Cackett2010,Disalvo2015} and can be described by a power-law with photon index ($\Gamma$) in the range 1.6--2.5, while the contribution from the thermal components ($ < 1$ keV) is significantly weakened \citep{Yoshida1993}.

Some LMXBs are known to exhibit periodic intensity dips in their X-ray light curves. Studies indicate that the intrinsic spectral properties of dippers are similar to that of non-dipping LMXBs with the explanation for intensity dips as the presence of ionized absorber in the external regions of accretion disc \citep{Boirin2005,Diaz2006}. The primary reason for the observation of more pronounced intensity dips in dippers is their higher edge-on inclination angle. Absorption features due to highly ionized species have been observed in the spectra of almost all high-inclination LMXBs in soft states \citep[e.g.,][]{Diaz2006, Neilsen2009, Ponti2012, Raman2018}.

\citet{Neilsen2009} and \citet{Ponti2012} inferred the presence of disc-wind outflows with large velocities in Black Hole (BH) LMXBs. Similar signatures of wind outflows have been reported in about 30 per cent of the NS LMXB population \citep{Diaz2016}. Moreover, the presence of these features in the high-inclination LMXBs indicate that the absorbers responsible for such features have an equatorial distribution. Many studies have been done to study the presence of spectral similarities between ordinary LMXBs, absorption features, disc winds and properties of Comptonizing corona \citep[e.g.,][] {Zhang2014, Zhang2016,Raman2018, Sharma2018,Gambino2019}. The present work focuses on the broad-band spectral study of one such dipping and eclipsing NS LMXB, XTE J1710-281.

XTE J1710-281 is a transient LMXB which is known to exhibit eclipses and intensity dips in its X-ray light curve \citep{Markwardt2001}. It was discovered with Rossi X-ray Timing Explorer (\emph{RXTE}) in 1998 \citep{Markwardt1998}. It has shown several Type-I thermonuclear X-ray bursts which indicates that the compact object in the system is a NS having an estimated source distance of about 15--20 kpc \citep{Markwardt2001}. It is a highly variable source with flux changes between 2 and 10 mCrab on time-scales of nearly 30 days \citep{Markwardt2001}. It shows intensity dips recurring at an orbital period of 3.28 h. The eclipses are known to last for an average duration of 420 s, excluding the ingress and egress phase of the eclipse \citep{Jain2011}. The source inclination is about $75^\circ - 80^\circ$ \citep{Younes2009}.

The primeval X-ray spectra of XTE J1710-281 determined with \emph{RXTE} data were described with either a thermal bremsstrahlung component ($kT$ = $14 \pm 3$ keV) or a power-law having photon index $1.8 \pm 0.1$ with column density $N{_{\rm H}\ < 2 \times 10^{22}}$ cm$^{-2}$ \citep{Markwardt1998}. 
\citet{Younes2009} studied the 0.2--10 keV spectra of this source by using the \emph{XMM-Newton} data and inferred that the spectra become harder during the dips. They modelled the changes between shallow and deep dipping spectra by using the approach of partial covering of power-law that resulted in the increase of covering fraction while neutral column density decreased from shallow dipping to deep dipping. In another approach, they used an ionized absorber model, and the derived results were consistent with those of several dippers observed with \emph{XMM-Newton}. However, some of the spectral parameters were not constrained for the persistent spectra. 

More recently, \citet{Raman2018} studied the 0.6--9 keV time-averaged and intensity resolved spectra of XTE J1710-281 with the \emph{Chandra} and the \emph{Suzaku} data. The spectra were well described with a power-law and partially ionized absorber. They detected Fe absorption lines at $\sim 6.60$ keV and $\sim 7.01$ keV along with a broad Fe emission line blend at 0.72 keV in the \emph{Suzaku} spectrum, a signature of accretion disc winds.

In this paper, we report the results of time-averaged broad-band spectral analysis of the persistent emission of XTE J1710-281 by using the \emph{Suzaku} archival data acquired in 2010 March (same as that analysed by \citet{Raman2018}, but covering energies above 10 keV for the first time). We have employed a source distance of 15 kpc \citep{Markwardt2001} and an inclination angle of ${80^\circ}$ \citep{Younes2009} throughout the paper.

\section{Observations and Data Analysis}
The Japanese space observatory \emph{Suzaku} \citep{Mitsuda2007} observed XTE J1710-281 starting at 22:05:54 on 2010 March 23 till 03:30:17 on 2010 March 26 in normal mode and collected data in $3 \times 3$ and $5 \times 5$ pixel modes with a total exposure of 76 ks (ObsID 404068010) by using the X-ray Imaging Spectrometer \citep[XIS;][]{Koyama2007} and the Hard X-ray Detector \citep[HXD;][]{Takahashi2007}. The XIS detectors consist of four units numbered from 0 to 3 and covers 0.2--12 keV energy range. HXD-PIN extends the energy range to about 60 keV. Of the four XIS units, three are front-illuminated (FI) CCDs (XIS0, XIS2 \& XIS3) and one is back-illuminated (BI) CCD (XIS1). Since its breakdown in 2006, XIS2 no longer collects useful data. Therefore, we have utilized the data from XIS0, XIS1 \& XIS3 detectors in our analysis. Due to the statistical limitations, we did not use the data above 9 keV for XIS and below 15 keV for HXD-PIN. 

\subsection{Data Processing}
\subsubsection{XIS Spectra}
\label{sec:XIS Spectra} 
We have performed the standard screening and filtering on the unfiltered XIS event files by using the high-level routine \textsc{aepipeline} provided with the \textsc{heasoft} version 6.27. We reprocessed the event files by using the calibration database (CALDB) version 20181010 for XIS. We used the \textsc{pileest} script to check for any pileup in the event files. It was found that the data were not piled-up during the observation. We combined the cleaned event files for both 3$\times$3 and 5$\times$5 pixel mode for each CCD, by using the \textsc{xselect} tool, individually, and generated good time intervals corresponding to out-of-burst phase, eclipses and dips. The source spectra for XIS detectors were extracted from a circular region of $3.5^\prime$ centred at the source. The background spectra were extracted from an annular region with inner radius $4^\prime$ and outer radius $6^\prime$, also centred at the source position. We produced the redistribution matrix files for each detector by using the ftool \textsc{xisrmfgen} and used \textsc{xissimarfgen} to generate the ancillary response files with source coordinates R.A. $= 257.55125^\circ$ and Dec. $= -28.13166^\circ$ \citep{Ebisawa2003}. We generated the combined spectrum for the two FI detectors by adding the individual source spectrum, background spectrum, redistribution file and response file of XIS0 and XIS3 detectors by using the \textsc{addascaspec} task. The net exposure time of the combined XIS0 and XIS3 spectrum was 48.8 ks and 24.4 ks for XIS1. We re-binned the final spectrum by using the ftool \textsc{grppha} to contain a minimum of 100 counts per bin and used $\chi^{2}$ statistics to check the goodness of fit. The XIS0--XIS3 combined spectrum was modelled in the energy range 0.6--9.0 keV. Due to the calibration uncertainties below 0.7 keV, we limited the lower energy bound to 0.7 keV for XIS1 spectrum. The channels between 1.7--2.3 keV were also removed to eliminate the artificial structures arising due to the Au and Si edges \citep{Zhang2014}.

\subsubsection{HXD-PIN Spectra}
 We have followed the standard reduction method for HXD-PIN spectrum as given by the \emph{Suzaku} team in the \emph{Suzaku} ABC Guide\footnote{https://heasarc.gsfc.nasa.gov/docs/suzaku/analysis/abc/abc.html} and have processed the unfiltered event files with HXD CALDB 20110913. We used the high-level task \textsc{hxdpinxbpi} to generate the HXD-PIN spectrum which corrects the spectrum for the dead time by using pseudo-events files available with the archived data and the same GTI file as used for XIS spectra. It also corrects the source spectrum for Non-X-ray Background (NXB) and Cosmic X-ray Background (CXB) by using the tuned PIN background event file distributed by the HXD team according to the time of observation. This task simulated and modelled the CXB contribution to the source spectrum using the model given by \citet{Boldt1987}.

\begin{figure}
    \includegraphics[width=\columnwidth]{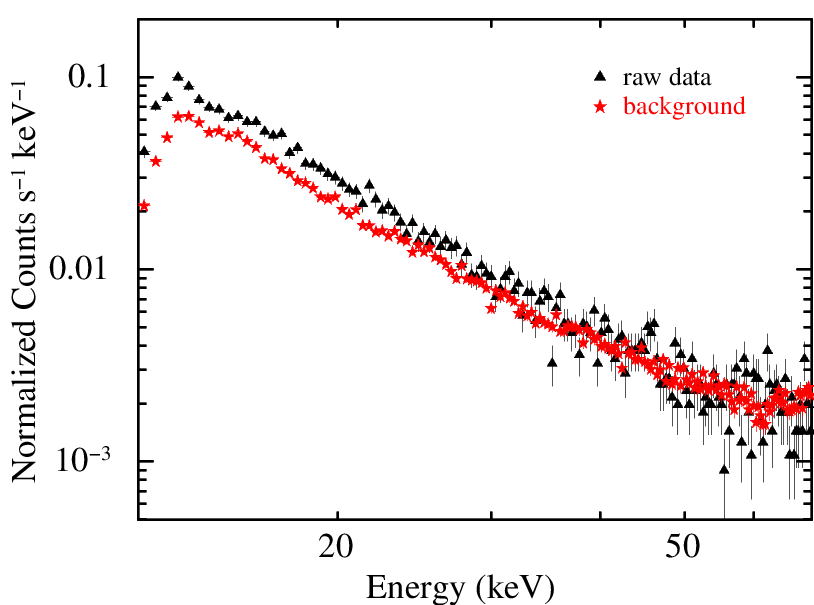}
    \caption{A comparison between the source and the background spectrum of XTE J1710 for the HXD-PIN data. The source was significantly detected above the background for energies up to 30 keV.}
    \label{fig:back}
\end{figure}

It finally updated the exposure time of the NXB spectrum by a factor of 10 as the NXB event file was created with 10 times the real background count rates for better statistics. The remaining exposure for the final spectrum after filtering was 14.8 ks. We used the \textit{ae\_hxd\_pinxinome8\_20100731.rsp} response file for HXD-PIN, provided by the HXD team according to the observation epoch. 

the energy range of HXD-PIN extends to about 60 keV, but as shown in Fig.~\ref{fig:back}, the source was detected above the background for energies up to 30 keV only. Therefore, we have limited our analysis to energies up to 30 keV only.

We used \textsc{xspec} \citep{Arnaud1996} version 12.11.0j and the models therein for the simultaneous spectral fitting of XIS and HXD-PIN spectra. In order to account for the different calibrations between the instruments, we included a cross-calibration factor in the spectral modelling. The value of the constant was fixed to 1 for the combined XIS0--XIS3 spectrum and left free for XIS1. The value for the HXD-PIN spectrum was fixed at 1.158 \citep{Kokubun2007}. We have used the updated photo-ionization cross-sections of \citet{Verner1996} and solar abundances by \citet{Wilms2000} to model the spectra throughout the paper.

\subsection{Light Curve}
\begin{figure}
    \includegraphics[width=\columnwidth]{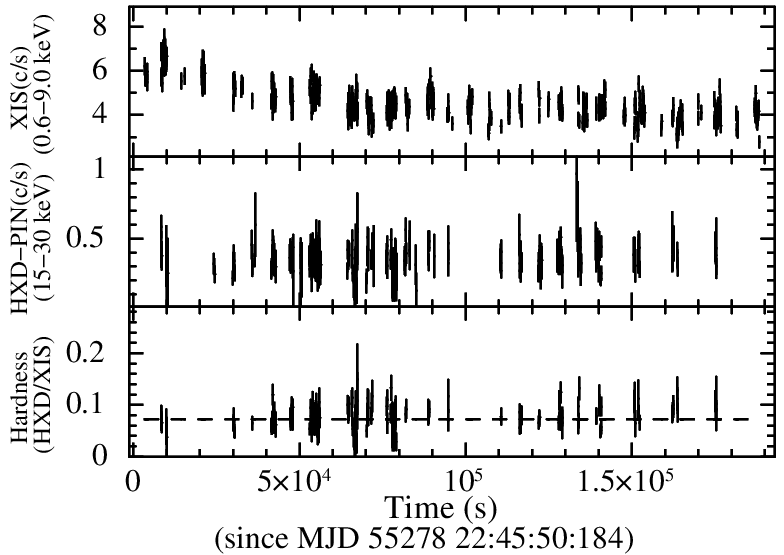}
    \caption{Light curve for XIS0 and HXD-PIN with 100 s binning. The top panel shows the persistent light curve for XTE J1710-281 from XIS0 in the energy range of 0.6--9.0 keV. Middle panel gives the HXD-PIN light curve in 15--30 keV and bottom panel is the hardness ratio defined as the ratio between HXD-PIN and XIS count rate.}
    \label{fig:light}
\end{figure}

The light curves along with the spectra for XIS0, XIS1, XIS3 and HXD-PIN were extracted by using the task \textsc{xselect}. We used the same region files and good time intervals as for the spectra to filter the light curves. The extracted light curves are shown in Fig.~\ref{fig:light}. We generated the XIS light curves in the energy range 0.6--9.0 keV with a binning of 100 seconds.

In Fig.~\ref{fig:light}, the top panel shows the clean background-subtracted light curve from XIS0. The raw light curve consisted of several dips, eclipse, six thermonuclear X-ray bursts and significantly large data gaps \citep{Raman2018}. We removed the dips, eclipses and bursts from the light curve. Initially, the XIS count rate increased slightly from an average of $\approx$ 6 counts s$^{-1}$ to $\approx$ 7 counts s$^{-1}$ and returned to $\approx$ 6 counts s$^{-1}$. It then gradually decreased through $\approx$ 5 counts s$^{-1}$ to $\approx$ 4 counts s$^{-1}$ fluctuating in-between.

In the middle panel, we have shown the HXD-PIN light curve for 15--30 keV energy range without any background correction. After filtering out the dips, eclipses and the bursts from the raw light curve, it was found that the HXD-PIN count rate initially remained constant at about 0.4 counts s$^{-1}$. There were negligible counts in HXD-PIN for the time when the XIS count rate decreased from 7 to 6 counts s$^{-1}$. It could be due to changes in the lower threshold level of PIN close to the observation date and good time filtering in \textsc{xselect}.

the last panel gives the hardness ratio defined as the ratio of HXD-PIN count rate to XIS count rate. The dashed horizontal line in this panel represents the weighted average of hardness ratio ($= 7.2 \pm 0.2\times 10^{-2}$). Although we observed a modest evolution of hardness ratio near the end of the exposure window, but due to the resolution limit of HXD-PIN data, we chose not to divide the observation further. As a result, the entire exposure has been used for the spectral modelling \citep[e.g.][]{Zhang2014}. 

\subsection{Spectral Analysis}
The reported spectral studies of the persistent emission of XTE J1710-281 agree with a model consisting of a power-law and an absorption component to account for the absorption due to the Inter-Stellar Medium (ISM). But, so far the spectra have been modelled up to 10 keV only \citep{Younes2009, Raman2018}. We have extended the spectral study of XTE J1710-281 to 30 keV.  Instead of using the power-law model as done in the reported works, we have incorporated the thermal Comptonization model \texttt{nthcomp} \citep{Zdziarski1996, Zycki1999} to model the broad-band time-averaged spectra of XTE J1710-281 as it facilitates in choosing the source of seed photons characterized by the parameter \textit{inp\_type} and has an asymptotic power-law index $\Gamma$.

\begin{figure}
   \includegraphics[width=\columnwidth]{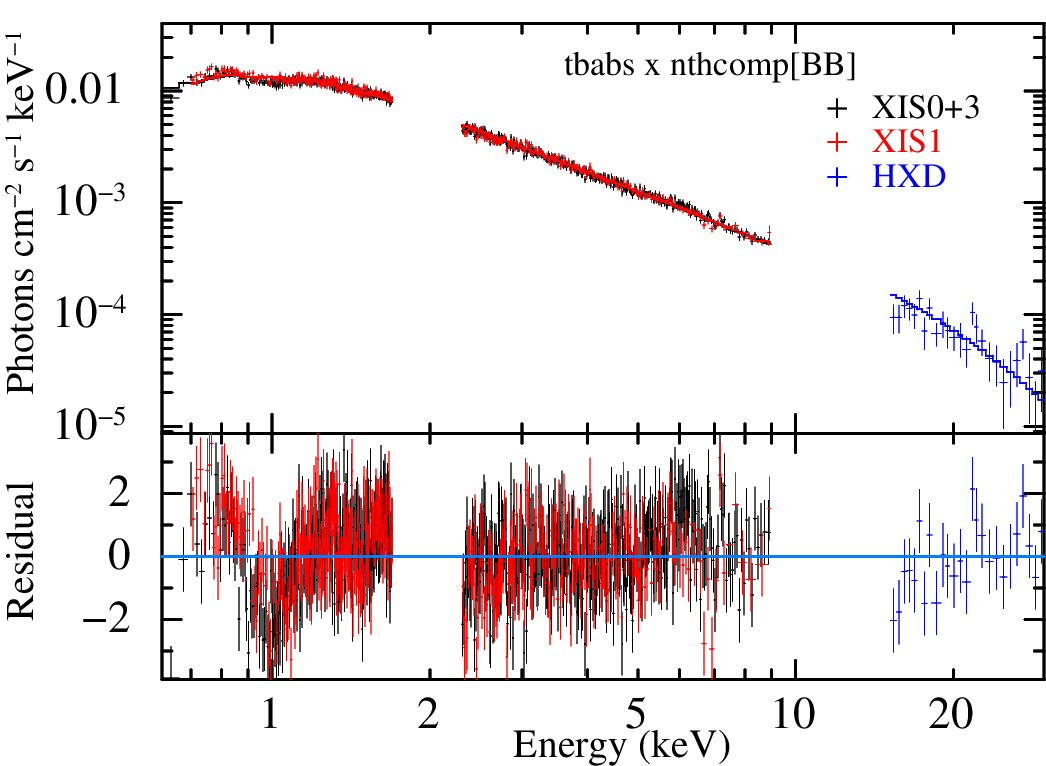}
    \caption{The spectrum of XTE J1710-281 modelled with \texttt{tbabs*nthcomp[BB]}. Large residuals below 2 keV and between 5--9 keV is evident.}
    \label{fig:3}
\end{figure}

We started with the simplest Comptonization model for modelling the XIS+PIN spectra. We used \texttt{tbabs}, to account for the absorption due to ISM in the direction of the line of sight of source, along with \texttt{nthcomp} by varying the seed photon source. Firstly, we assumed the source of seed photons as the NS surface/boundary layer (\texttt{nthcomp[BB]}) and then followed with the seed photon source as accretion disc (\texttt{nthcomp[diskbb]}). The resulting fits gave poor results with a $\chi^{2}$ of 2081 and 2099 for 1761 degrees of freedom (dof), respectively, along with large residuals below 2 keV and between 5 to 9 keV (Fig.~\ref{fig:3}).

In order to account for the large residuals at lower energy, we added either (i) \texttt{bbodyrad}, a single temperature blackbody model that accounts for the contribution from the NS surface/boundary layer emission or (ii) \texttt{diskbb} that models the emission from the accretion disc. The \texttt{bbodyrad} component is described by two parameters, temperature of the body $kT_{\rm BB}$ and normalization $N_{\rm BB}$ related to the radius of the emission region as $N_{\rm BB}=(R/D_{\rm 10})^{2}$, where $R$ is the radius in km and $D_{\rm 10}$ is the source distance in units of 10 kpc. The other soft component \texttt{diskbb} consists of two parameters, temperature $kT_{\rm disc}$ at the inner disc edge and normalization $N_{\rm disc}=(R_{\rm in}/D_{\rm 10})^{2} {\rm {cos}\theta}$, where $R_{\rm in}$ is the apparent radius of the inner disc, $D_{\rm 10}$ being the source distance in units of 10 kpc, and $\theta$ is inclination of accretion disc. The addition of \texttt{bbodyrad} or \texttt{diskbb} to the Comptonization component (\texttt{nthcomp}) with either seed sources improved the fit statistically by reducing the $\chi^{2}$ to some extent, but the fits gave low values of $kT_{\rm seed}$ ($\sim$ 0.22 keV) and large emission radius ($>$ 20 km) for the case of blackbody seed. Also, the residuals near 0.7 keV and 6 keV were not modelled. The feature at 6 keV mimicked an emission feature. We added two Gaussian features to model the residuals at 0.60 keV and 5.98 keV. However, the addition of Gaussian component at 5.98 keV was not statistically significant as it improved $\chi^2$ by 18 for 3 $\Delta$dof. Thus, the final model that provided the best-fitting with physically acceptable parameters was \texttt{tbabs*(diskbb + nthcomp[diskbb] + gaus + gaus)} (M1, hereafter). We have used this model as the base model in our study. 

As we chose the accretion disc as the seed photon source, we tied the seed photon temperature ($kT_{\rm seed}$) to the disc temperature ($kT_{\rm disc}$) and the resulting fit gave a ${\chi^{2}}$ of 1767 for 1754 dof with a foreground column density of $6.8_{-1.0}^{+1.2}\times 10^{21}$ cm$^{-2}$, disc temperature $kT_{\rm disc} = 0.28_{-0.02}^{+0.03}$ keV, photon index $\Gamma = 1.86 \pm 0.02$ and electron corona temperature $kT_{\rm e} = 5.15_{-0.49}^{+0.61}$ keV. Using the corona temperature and photon index we have calculated the optical depth of corona, $\tau$, by employing the equation from \citet{Zdziarski1996}, 
\begin{equation}
 \Gamma = \sqrt{\frac{9}{4} + \frac{1}{\frac{kT_{\rm e}}{m_{\rm e}c^{2}}\ \tau \left(1 + \frac{\tau}{3}\right)}} - \frac{1}{2}
\label{eq:tau}
\end{equation}
where $m_{\rm e}$ is the rest mass of the electron. We estimated the optical depth to be $8.09_{-0.63}^{+0.62}$ for this model which is consistent with the results of \citet{Lin2007} for LMXBs.

Then, in order to check the certainty of source seed, we estimated the seed photon emission region size using the Eqn.~\ref{eq:radius}, which presumes a spherical geometry for emission region, \citep{intzand1999}. 
\begin{equation}
 R_{\rm o}\ =\ 3\times 10^{4}\ D_{\rm kpc} \left[\frac{f_{\rm bol}}{(1+y)}\right]^{1/2}\ (kT_{\rm seed})^{-2}\ \rm km
\label{eq:radius}
\end{equation}
where $D_{\rm kpc}$ is the source distance in kpc, $f_{\rm bol}$ is the unabsorbed bolometric flux for Comptonization in the energy range of 0.1--100 keV and $y= 4kT_{\rm e}\tau^{2}/(m_{\rm e}c^{2})$ is the Comptonization parameter that gives the relative energy gain of photons. The value of emission radius $R_{\rm o} = 42 \pm 11$ km implies that the emission arises well away from the compact NS surface/boundary layer, thereby justifying the inner disc Comptonization \citep{Sharma2018}. Table~\ref{tab:bestfit} reports the best-fitting parameters for this model and all the subsequent models. It was observed that the model with blackbody as the Comptonization seed was not consistent. When we set the seed source as NS surface/boundary layer, the fit gave an unreasonably small value of $kT_{\rm seed} = 0.22$ keV which implies a large emission radius ($> 20$ km) which contradicts the fact that seed is provided by the NS itself.
\begin{figure*}
   \includegraphics[width=\columnwidth]{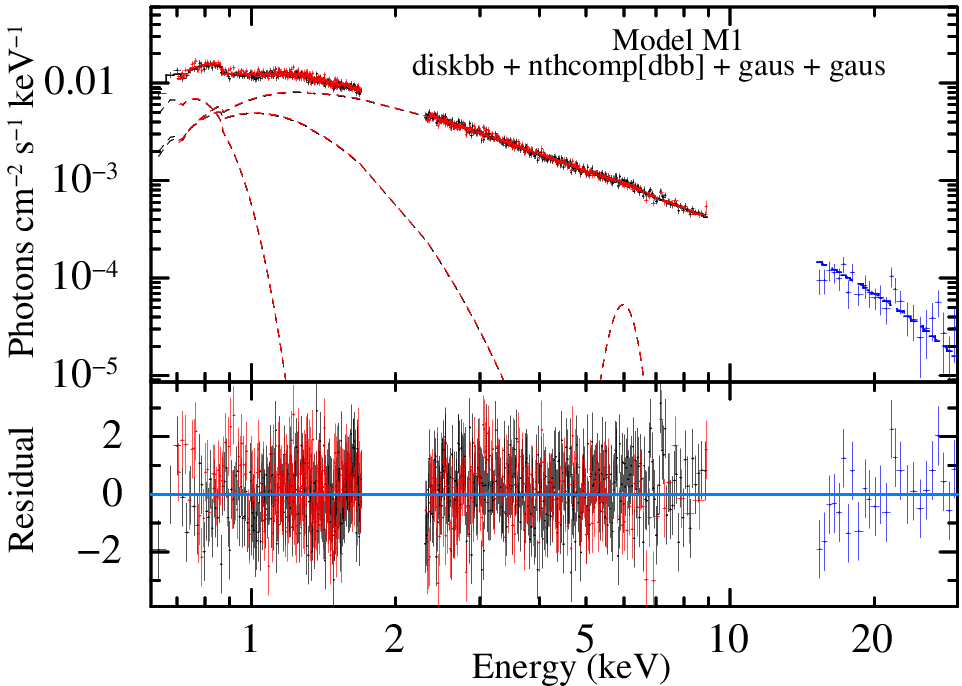}
   \includegraphics[width=\columnwidth]{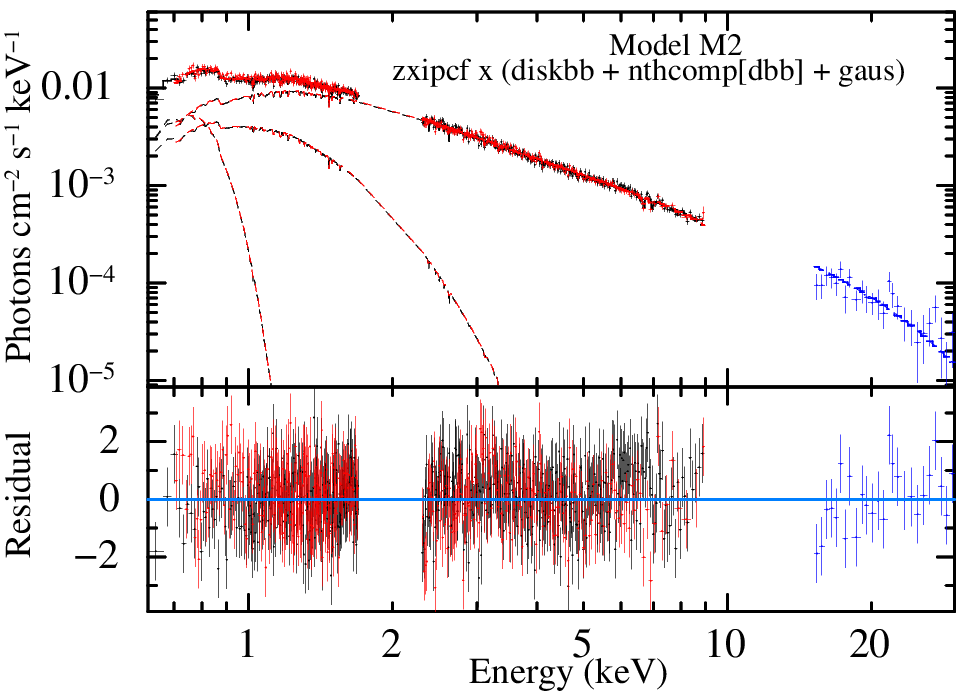}
 \centering
   \includegraphics[width=\columnwidth]{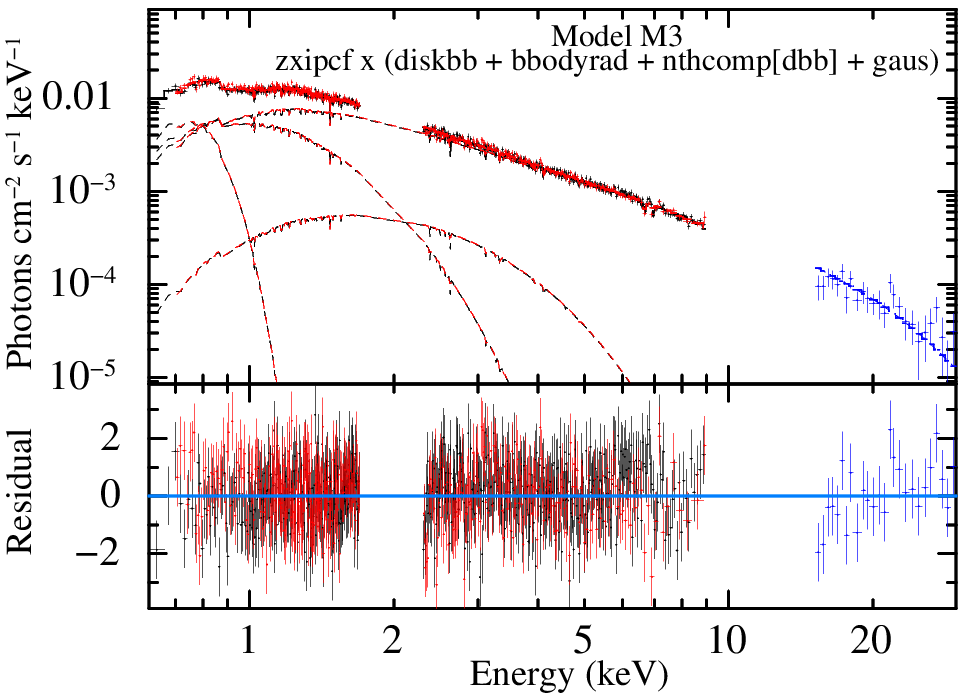}
\caption{best-fitting spectra for the time-average persistent XIS (0.6--9.0 keV) and HXD-PIN (15--30 keV) spectra of XTE J1710-281. the upper panel in each plot gives the best-fitting unfolded model and spectrum, and the bottom panel represents the residuals with respect to the model.}
 \label{fig:spec}
\end{figure*}

The XIS light curve of XTE J1710-281 (Fig.~\ref{fig:light}) shows a decaying trend for about 50 ks of the observation and then the count rate becomes almost constant. We extracted the XIS spectra from these two segments of the light curve to check the consistency of the spectral model. We did not find any significant difference between the two spectra. It implies that there is no significant change in the intrinsic emission spectrum of the source.

 In order to compare our results with the known published works, we have included a partial covering ionized absorber component \texttt{zxipcf} \citep{Reeves2008}. XTE J1710-281 exhibits intensity dips. Therefore, \texttt{zxipcf} will account for the presence of local ionized absorber around the NS which is an ubiquitous feature of LMXBs primarily detected in high-inclination sources \citep{Diaz2006,Diaz2016,Sharma2018,Iaria2019}. The ionized absorber column density $N_{\rm H}^{\rm ion}$, ionization parameter log($\xi$) and covered source fraction $Cvr.f$ characterizes the \texttt{zxipcf} component. It modelled the broad feature near 6 keV to some extent and improved the $\chi^{2}$ to 1752 for same 1754 dof. The emission feature at 0.6 keV was modelled with the Gaussian component. Thus, the model \texttt{tbabs*zxipcf*(diskbb + nthcomp[diskbb] + gaus)} (M2, hereafter) gave an efficient and acceptable fit for the spectra with $kT_{\rm disc} = kT_{\rm seed} = 0.30_{-0.03}^{+0.04}$ keV, optical depth $\tau = 8.35 \pm 0.57$, emission region size of seed photons $R_{\rm o}= 37_{-13}^{+11}$ km, ionized column density of $2.32\times 10^{24}$ cm$^{-2}$, ionization parameter $10^{4.05}$ erg cm s$^{-1}$ and a covering fraction of 0.33 (Table~\ref{tab:bestfit}). We computed the 0.1--100 keV unabsorbed bolometric flux for individual components as well as for the entire model by using the convolving model \texttt{cflux}. With respect to M1, there was a considerable decrease in the bolometric flux from the disc and an increment in the Comptonization flux, for M2. The fractional share from disc and Comptonization to the total unabsorbed flux changed from 0.23 to 0.16 and 0.77 to 0.84 respectively, for M2.

When we changed the soft component to blackbody, we obtained the best-fitting with a $\chi^{2}/{\rm dof}=1751/1753$ and $kT_{\rm BB} = 0.24$ keV. The spectral parameters were somewhat identical as the previous model M2, and it served as the identical model with only the difference of soft component, and a large blackbody radius $> 35$ km, thus, we discarded this model. Also, with the seed source as NS surface/boundary layer, both of these models did not provide satisfactory fits in terms of spectral parameter values. 

Finally, we tried to model the persistent spectra with two soft components and one hard Comptonization component \citep{Lin2007, Armas2017} along with the partially ionized absorber. We used \texttt{bbodyrad}, \texttt{diskbb} and \texttt{nthcomp} to model the spectra. Again, we modelled the broad emission feature at 0.6 keV using the Gaussian component. The model gave a satisfactory fit with either blackbody or disc as a source for seed photons with similar spectral parameters. For the sake of consistency, we stick with the disc seed photon. The resulting model (M3, hereafter) explained the spectra well with reasonable spectral parameter values (Table~\ref{tab:bestfit}) and a $\chi^{2}/{\rm dof}$ of 1749.6/1752. While most of the best-fitting parameters of M3 were consistent with the previous models M1 and M2, there was a marginal increment in the optical depth $\tau = 9.5$ and slight decrement in the photon index to 1.77 for this model. 

Fig.~\ref{fig:spec} shows the best-fitting spectra along with the residuals for all the three models discussed above. The spectral fits indicate the presence of positive residuals around 7.5 keV for model M1. \texttt{zxipcf} modelled the 6.0 keV feature for M2 and M3. We tried to remove the feature by adding Gaussian component around 7.5 keV but, this improvement did not prove to be statistically significant as it reduced the $\chi^{2}$ by 13 for a $\Delta$dof of 3 for M1. We, therefore, did not include this component in the model and our analysis. 

Following the works of \citet{Zhang2014} and \citet{Gambino2019}, where the spectra of dipping LMXBs were modelled with a double seed model, we also tried to model the spectra of XTE J1710-281 by using two Comptonization components where one component has seed photon distribution governed by the blackbody while the other component is fed with photons distributed according to multi-colour disc blackbody. We also tied the corresponding seed photon temperatures with blackbody and disc blackbody components. Further, with the assumption that the Comptonization taking place is originating from the same corona, we linked the $\Gamma$ and $kT_{\rm e}$ across the two components. This model provided a statistically good fit ($\chi^2/{\rm dof} = 1778.2/1754$) to the spectra, but there was no improvement compared to the previous models. Also, the value of blackbody normalization drifted to a very small value ($< 0.1$) to be physical while $kT_{\rm e}$ could not be constrained, thus we discarded this model.

We did not detect any absorption lines due to the highly ionized elements at 6.6 keV and 7 keV in the persistent spectra, which is considered a signature of dipping LMXBs. \citet{Raman2018} reported the presence of these lines in the entire time-averaged spectra. We also detected the presence of one of the two lines in the average spectra, cleaned only for bursts, at 6.59 keV. The absence of this line in the persistent spectra and its presence in the average spectra implies a strong correlation between the origin of such features and the absorber responsible for the dips.

\begin{table*}
\caption{best-fitting spectral parameters for broad-band time-averaged \emph{Suzaku} spectra for XIS and HXD-PIN. The errors for each parameter are quoted at 90\% confidence level.}
\label{tab:bestfit}
\resizebox{0.80\linewidth}{!}{
\begin{tabular}{c c c c c} 
\hline
\textbf{Component} & \textbf{Parameters} & \textbf{Model M1} & \textbf{Model M2} & \textbf{Model M3} \\
\hline

TBABS & $N_{\rm H}$ ($10^{22}$ cm$^{-2}$) &$0.68_{-0.10}^{+0.12}$ & $0.55_{-0.12}^{+0.13} $ & $0.59_{-0.14}^{+0.18} $\\[0.7ex]

 ZXIPCF & $N_{\rm H}^{\rm ion}$ ($10^{22}$ cm$^{-2}$) & & $232_{-142}^{+259} $ & $243_{-88}^{+243} $ \\[0.5ex]
  & log($\xi$) & & $ 4.05_{-0.31}^{+0.18} $ & $4.10_{-0.25}^{+0.16} $ \\[0.5ex]
  & $Cvr. f$ & & $0.33_{-0.12}^{+0.16} $  & $0.38_{-0.13}^{+0.20} $ \\[0.7ex]

 DISKBB & $kT_{\rm disc}$ (keV) & $0.28_{-0.02}^{+0.03} $ & $0.30_{-0.03}^{+0.04} $ & $0.30_{-0.04}^{+0.06} $ \\ [0.5ex]
 & $N_{\rm disc}$ & $437_{-225}^{+457}$ & $230_{-155}^{+363} $ & $333_{-203}^{+628} $ \\[0.5ex]
 & $R_{\rm disc}$ (km) & $75_{-23}^{+32} $ & $55_{-23}^{+33} $ & $ 66_{-25}^{+45} $ \\[0.5ex]
 & $f^{\dag}_{\rm bol}$ ($10^{-10}$ erg cm$^{-2}$ s$^{-1}$) & $0.57 \pm 0.02
 $ & $ 0.37 \pm 0.01 $ & $ 0.53 \pm 0.01 $ \\[0.7ex]

 BBODYRAD & $kT_{\rm BB}$ (keV) & & & $0.65_{-0.53}^{+0.26} $ \\[0.5ex]
 & $N_{\rm BB}$ & & & $ < 7 $ \\[0.5ex]
 & $R_{\rm BB}$ (km) & & & $ < 4 $ \\[0.5ex]
 & $f^{\dag}_{\rm bol}$ ($10^{-10}$ erg cm$^{-2}$ s$^{-1}$) & & & $0.06 \pm 0.01 $ \\[0.7ex]

 NTHCOMP & $\Gamma$ & $1.86 \pm 0.02 $ & $ 1.85 \pm 0.02$ & $1.77_{-0.06}^{+0.08} $ \\[0.5ex]
 & $kT_{\rm e}$ (keV) & $ 5.15_{-0.49}^{+0.61} $ & $4.94_{-0.41}^{+0.49} $ & $4.45_{-0.55}^{+0.69} $\\[0.5ex]
 & $kT_{\rm seed}$ (keV) & $=kT_{\rm disc}$ &$=kT_{\rm disc}$ & $=kT_{\rm disc}$ \\[0.5ex]
 & $inp\_type$ & 1.0 & 1.0 & 1.0\\[0.5ex]
 & Norm ($10^{-2}$) & $2.20 \pm 0.14 $ & $2.24_{-0.16}^{+0.18} $ & $1.91_{-0.54}^{+0.41} $ \\[0.5ex]
 & $\tau$ & $8.09_{-0.63}^{+0.62}$ & $ 8.35 \pm 0.57$ & $9.50_{-1.33}^{+1.31}$ \\[0.5ex]
 & $y$ & $2.64_{-0.61}^{+0.78}$ & $2.70_{-0.55}^{+0.69}$ & $3.14_{-1.11}^{+1.56}$ \\[0.5ex]
 & $R_{\rm o}$ (km)& $42 \pm 11$ & $ 37_{-11}^{+13}$ & $ 34_{-14}^{+19}$ \\[0.5ex]
 & $f^{\S}_{\rm bol}$ ($10^{-10}$ erg cm$^{-2}$ s$^{-1}$) &$1.94 \pm 0.01$ & $ 2.01 \pm 0.02 $ & $1.86 \pm 0.06 $ \\[0.7ex]

 GAUS & $E_{\rm line}$ (keV) & $0.60_{-0.09}^{+0.07} $ & $0.67_{-0.11}^{+0.06}$ & $0.64_{-0.14}^{+0.07} $ \\[0.5ex]
 & $\sigma$ (keV) & $ 0.14_{-0.03}^{+0.04} $ & $0.11_{-0.03}^{+0.04} $ & $0.12_{-0.03}^{+0.05} $ \\[0.5ex]
 & Norm ($10^{-2}$) & $4.10_{-2.53}^{+15.35} $ &$1.21_{-0.83}^{+3.60}$ & $1.79_{-1.27}^{+8.36} $ \\[0.5ex]
 &  EW (eV) &$364_{-364}^{+72}$ & $191_{-191}^{+73} $ & $208_{-208}^{+160} $ \\[0.7ex]

 GAUS & $E_{\rm line}$ (keV) & $5.98_{-0.18}^{+0.16} $ & &  \\[0.5ex]
 & $\sigma$ (keV) & $0.33_{-0.16}^{+0.20} $ & & \\[0.5ex]
 & Norm ($10^{-5}$) & $ 4.44_{-2.09}^{+2.62}$ & & \\[0.5ex]
 & EW (eV) &$49_{-27}^{+28}$ & & \\[0.7ex]

 CONS & $C_{\rm XIS1}$ &$ 1.01 \pm 0.01$ & $ 1.01 \pm 0.01$ &$ 1.01 \pm 0.01$ \\[0.5ex]
 & & & & \\
 &$f^{\ddag}_{\rm Total}$ ( $10^{-10}$ erg cm$^{-2}$ s$^{-1}$) &$ 2.51 \pm 0.01 $ & $2.38 \pm 0.01  $ & $2.46 \pm 0.01  $ \\[0.5ex]
&$L^{*}_{\rm X}$ ( $10^{36}$ erg s$^{-1}$) &$ 6.76 \pm 0.03 $ & $ 6.41 \pm 0.03 $ & $ 6.63 \pm 0.03 $ \\[0.7ex]
\hline
 & $\chi^2$/dof & 1766.9/1754 & 1752.1/1754 & 1749.6/1752 \\
\hline

 \multicolumn{5}{l}{Model M1 = \texttt{tbabs*(nthcomp[diskbb] + diskbb + gaus + gaus)}} \\
 \multicolumn{5}{l}{Model M2 = \texttt{tbabs*zxipcf*(nthcomp[diskbb] + diskbb + gaus)}} \\
  \multicolumn{5}{l}{Model M3 = \texttt{tbabs*zxipcf*(nthcomp[diskbb] + bbodyrad + diskbb + gaus)}} \\
 \multicolumn{5}{l}{$^{\dag} f_{\rm bol}$ is the unabsorbed bolometric flux in 0.1--100 keV without Comptonization contribution.} \\
 
 \multicolumn{5}{l}{$^{\S} f_{\rm bol}$ is the unabsorbed bolometric flux due to Comptonization only in 0.1--100 keV.} \\
 \multicolumn{5}{l}{$^{\ddag} f_{\rm Total}$ is the total unabsorbed bolometric flux in 0.1--100 keV.} \\
 \multicolumn{5}{l}{$^{*} L_{\rm X}$ is the unabsorbed 0.1--100 keV X-ray luminosity.} \\
 \multicolumn{5}{l}{Parameters are calculated for an assumed distance of 15 kpc and an inclination of ${80^\circ}$.} \\
  \multicolumn{5}{l}{$C_{\rm XIS1}$ represents the cross-calibration factor for XIS1 with respect to the combined XIS0--XIS3.}
\end{tabular}}
\end{table*}


\section{Discussion and Conclusion}
We have performed and presented a detailed spectral analysis of the time-averaged persistent emission spectra of LMXB XTE J1710-281, by using the archival data collected by the Japanese space mission \emph{Suzaku}. We have utilized the XIS and HXD-PIN data for extending the spectral analysis up to 30 keV. We have also presented the light curve for the source from both XIS and HXD-PIN after removing the dips, bursts and eclipses. We present a much clearer and complete picture of the emission process and surroundings of the system. The earlier studies using \emph{XMM-Newton}, \emph{Chandra} and \emph{Suzaku} data explored the spectra up to 10 keV only \citep{Younes2009,Raman2018} and could not quantify the emission components, origin and contribution of different components, temperature of the emitting region and surrounding environment. We have incorporated a modelling scheme to quantify these properties and provide a more comprehensible description of the source.

We have used the typical model comprising a single temperature blackbody and/or a multi-colour accretion disc component along with a hard Comptonization component \citep{Lin2007}. We successfully modelled the broad-band time-averaged persistent spectra from \emph{Suzaku} observation with three models. We used the combination of a single thermal component with accretion disc Comptonization as the first model that gave a low disc temperature of $0.28_{-0.02}^{+0.03}$ keV and a moderate corona temperature $kT_{\rm e} = 5.15$ keV.

We have modelled the spectra with partially covering ionized absorber (\texttt{zxipcf}) to account for the highly ionized material obscuring the system and obtained very well constrained results for the spectral parameters. The foreground column density changed slightly, but the disc temperature and seed temperature remained significantly consistent across the three models. While the spectral parameters were consistent across the models, the inclusion of \texttt{zxipcf} provided a better fit in terms of $\chi^2$ and modelling of the feature near 6 keV. The ionized absorber column density turned out to be $2.32\times 10^{24}$ and $2.43\times 10^{24}$ cm$^{-2}$, with an identical ionization parameter $\sim 10^{4}$ erg cm s$^{-1}$ for M2 and M3, respectively, implying a significantly high-ionized absorber density compared to $45.96\times 10^{22}$ cm$^{-2}$ and increased ionization by two orders of magnitude compared to the values reported by \citet{Raman2018} for their analysis of \emph{Suzaku} time-averaged spectrum. However, large errors in the value of ionized absorber column density makes it difficult to put stringent limits on it. The fits returned a similar covering fraction value of $\approx$ 0.35 for both the models.

We have modelled the spectra with two thermal component and an ionized absorber approach also (M3) and obtained the blackbody component temperature $kT_{\rm BB} \approx 0.6$ keV, which is well below the upper limit ($\sim$ 1 keV) for LMXBs in hard state \citep{Church2001,Barret2003}. The resulting $\chi^{2}$ accompanied with physically justified spectral parameters backed up the notion of contribution from the NS itself. However, the flux contribution from it remained significantly low at $\approx$ 2.4 per cent to the total unabsorbed bolometric flux while the contribution from disc increased to $\sim$ 22 per cent and that of Comptonization reduced by $\sim$ 8 per cent, compared to the model M2. For the blackbody component, we have computed the emission region radius using the normalization, to be $R_{\rm BB} < 4$ km, again consistent with the fact that the contribution was provided by the NS, itself, and implies that the emission originates from the equatorial region of the NS surface/boundary layer \citep{Zhang2014}.

We have obtained the 0.1-100 keV unabsorbed X-ray luminosity (without any correction), by assuming an isotropic emission and a source distance of 15 kpc \citep{Markwardt2001}, for all the three models. We have calculated the 0.1--100 keV bolometric luminosity as $6.76 \pm 0.03\times 10^{36}$, $6.41 \pm 0.03\times 10^{36}$ and $6.63 \pm 0.03\times 10^{36}$ erg s$^{-1}$ for M1, M2 and M3, respectively. This scales up to $\approx$ 0.04 L$_{\rm Edd}$ for M1, M2 and M3, for a NS of mass 1.4 M$_{\sun}$, as expected for the low/hard state of LMXBs \citep{vander2006}.

A thorough study of the spectral shape of XTE J1710-281 shows that the spectra have a flat, long-tail at higher energies. This pattern is similar to the one exhibited by other dipping LMXBs in the hard state and opposite to that of dipping LMXBs in the soft state (which show a steep tail) \citep{Zhang2014,Sharma2018,Gambino2019}. The photon index ($\Gamma \approx 1.85$) is consistent across the three models and with LMXBs in the hard spectral state \citep{Yoshida1993}. The value of photon index is in coherence with the reported values ($1.91 \pm 0.02$) for \emph{XMM-Newton} spectrum \citep{Younes2009} and ($1.79 \pm 0.02$) for intensity-resolved persistent spectrum from \emph{Suzaku} \citep{Raman2018}.

We have found a high value for the optical depth (an average of $\approx 8.5$ from the three models) of the corona. The optical depth is marginally higher than the suggested limit ($\sim 2-4$) for LMXBs in the hard state \citep{Barret2000}. The obtained values for the corona temperature, $kT_{\rm e} = 5.15_{-0.49}^{+0.61}$, $4.94_{-0.41}^{+0.49}$ and $4.45_{-0.55}^{+0.69}$ keV for M1, M2 and M3, respectively, are significantly greater than the values reported for other dippers (such as, 4U 1915-05, MXB 1658-298, XB 1254-690) observed in the soft state \citep{Iaria2001, Zhang2014, Sharma2018, Gambino2019}. However, it is quite lower than that reported for the MXB 1658-298 ($\sim$ 18 keV) and other LMXBs in the hard state \citep{Lin2007, Sharma2018}, with a possible reason of difference in the modelling scheme. Compared to the spectral parameters of accretion-powered millisecond X-ray pulsars (AMXPs), there is a significantly large difference in the corona temperature for the hard spectral state ($> 30$ keV) \citep[e.g., XTE J1751-305, IGR J17511-3057, IGR J16597-3704;][]{Gierlinski2005,Papitto2010,Sanna2018}. However, it is moderately higher than that observed for SAX J1748.9-2021 in the soft state \citep[$< 2.5$ keV;][]{Pintore2016,Sharma2019}. A moderate corona temperature accompanied by the high value of optical depth may imply that the source was in a hard or rather an intermediate state.

The degree of Comptonization for the first two models (M1 and M2) was evaluated to be $y = 2.64_{-0.61}^{+0.78}$ and $y = 2.70_{-0.55}^{+0.69}$ considerably greater than that for a soft spectral state and more close to the value reported for MXB 1658-298 ($y \sim$ 2.5) in hard state \citep{Sharma2018}. For M3, the Comptonization parameter turned out to be slightly higher at  $y = 3.14_{-1.11}^{+1.56}$. The high value of $y$ implies a stronger Compton up-scattering of the low-temperature seed photons ($kT_{\rm seed} \approx 0.30$ keV). Seed photons are scattered from the inner accretion disc region as they tend to pass through the significantly hot electron corona for a longer effective path, which in-turn explains the high value of $y$ \citep{Zhang2014, Zhang2016}.

 In our analysis, we realised that when seed photon source set to the NS surface/boundary layer (blackbody), then the model failed to provide an acceptable fit and returned a small seed photon temperature, resulting in an unlikely large emission radius while the model with disc as the seed source prevailed with physically acceptable spectral parameter values. We estimated the emission radii for Comptonization seed photon to be $\approx$ 42, 37 and 34 km for model M1, M2 and M3, respectively. Likewise, the inner disc radii calculated from the normalization parameter of \texttt{diskbb} were consistent with these values (Table~\ref{tab:bestfit}). It implied, we can assert that the inner accretion disc was the dominant source of seed photons for the Comptonization.

The value of the inner disc radius suggests that the disc might be truncated at a moderately farther distance from the NS. However, the failure in the detection of any pulsation from the source so far, signifies that the magnetosphere around the NS is either absent or very weak if present, like most accreting NS LMXBs \citep{Patruno2018}. Our limits on the inner disc radii ($R_{\rm disc}$) may imply the truncation of the accretion disc due to the magnetic pressure exerted by the magnetosphere of the NS \citep{Cackett2009,Ludlam2016}. Assuming that the accretion disc is truncated at the magnetospheric radius, we can estimate the limit on the magnetic field strength of the NS using the extreme limits on disc radius \citep{Degenaar2014,Sharma2019}. We have employed the modified Eqn.~\ref{eq:mag} to estimate the magnetic field strength (B) originally given by \citet{Ibragimov2009}:
\begin{equation}
 \begin{split}
    B\ =\ 2.4\times 10^{7}\ k_{\rm A}^{-7/2} \left(\frac{M}{1.4 \rm {M}_{\sun}}\right)^{1/4}\ \left(\frac{R_{\rm in}}{10\ \rm km}\right)^{7/4}\ \left(\frac{R}{10^{6}\ \rm cm}\right)^{-3}\\
 \times \left(\frac{f_{\rm ang}}{\eta}\ \frac{F_{\rm bol}}{10^{-9} \rm erg\ \rm cm^{-2}\ \rm s^{-1}}\right)^{1/2}\ \frac{D}{15\ \rm kpc}\ \ \rm G
\label{eq:mag}
 \end{split}
 \end{equation}
 In the equation, $k_{\rm A}$ is the geometry coefficient with a permissible value $\simeq$ 0 .5--1.1 \citep{Psaltis1999,Long2005,Kluz2007}, $f_{\rm ang}$ is the anisotropy correction factor with a value close to unity \citep{Ibragimov2009}, and $\eta$ is the accretion efficiency factor, generally taken to be $\approx$ 0.1. We have assumed $k_{\rm A}=1$, $\eta=0.1$, $f_{\rm ang}=1$, $D=15$ kpc, $R=10$ km and a reasonable mass of 1.4 M$_{\sun}$ along with the average 0.1--100 keV bolometric flux of $2.4\times 10^{-10}$ ergs cm$^{-2}$ s$^{-1}$ from our best-fitting results. We estimated the magnetic field strength to lie between $2.7\times 10^{8} < B < 2.5\times 10^{9}$ G, for the disc truncating between $31 < R_{\rm in} < 111$ km. This limit on the magnetic field is consistent with the typical limit of magnetic field for the NS LMXBs ($B \lesssim 10^{9}$ G) and AMXPs \citep{Cackett2009, Mukherjee2015, Ludlam2016, Sharma2020}.
 
We have detected a broad emission feature around 0.6 keV in the persistent spectra. However the flux of emission feature at 0.6 keV could not be constrained, due to the poor statistics at lower energies and we have found upper limits on EW of 436, 264 and 368 eV for M1, M2 and M3, respectively. \citet{Raman2018} reported the detection of a broad emission feature at 0.72 keV in the time-averaged \emph{Suzaku} spectrum while due to the flux limitation below 1 keV, they failed to detect it in the \emph{Chandra} spectrum. Similarly, \citet{Younes2009} did not detect any emission/absorption line in their spectral analysis of XTE J1710-281 with the \emph{XMM-Newton}. Detection of O \textsc{vii} and O \textsc{viii} emission lines at 0.6 keV have been reported in spectra of LMXBs such as EXO 0748-676 \citep{Cottam2001,Peet2009,Psaradaki2018}. But, due to the poor statistics in case of XTE J1710-281 and the fact that XIS detectors often suffer from contamination effects below 2 keV (with oxygen being one of the contaminants), it is difficult to comment whether this feature is an inherent effect due to contamination of XIS detectors or it is due to the blend of K$\alpha$ transitions of the ionized O \textsc{vii} -- O \textsc{viii} species. We propose that future observations with sensitive soft X-ray detectors to be scheduled to confirm the presence/absence of this feature in the persistent spectra of the source.

We also modelled the broad emission like feature at 5.98 keV, with an EW of $75_{-40}^{+42}$ eV for M1 while ionized absorber modelled this feature for M2 and M3. We found that this feature was not very statistically significant, yet we retained it in the model to suppress the feature in the residuals. We propose that this feature may be present due to our model configuration as this feature was not present in the analysis by \citet{Raman2018} and associating it with the neutral Fe line at 6.4 keV implied an unlikely high redshift, not very typical of NS LMXBs.

\citet{Raman2018} also reported the detection of an absorption feature at 6.6 keV associated with the blend of highly ionized Fe \textsc{xix} -- Fe \textsc{xxv} transitions and a weak absorption feature around 7.01 keV in the time-averaged spectra. We did not detect any absorption feature from highly ionized species in the average persistent spectra during our analysis. However, we found upper limits on the EW for 6.7 keV (Fe \textsc{xxv}) and 6.96 keV (Fe \textsc{xxvi}) lines by fitting Gaussian profiles at these energies and fixing width at 20 eV \citep{Gambino2019}. We derived an upper limit of 23 eV for 6.7 keV line and 21 eV for 6.9 keV line with the neutral absorber. We also modelled an absorption feature at $6.59 \pm 0.03$ keV with EW of $29_{-5}^{+77}$ eV and derived the upper limit of 15 eV for 6.9 keV line in the time-average XIS0--XIS3 spectrum. While these absorption features were present in the averaged spectra, the persistent spectra without dips, bursts and eclipses did not show their presence. One possible reason can be that these features are the signature of the absorbing atmosphere surrounding the system responsible for the intensity dips. Once we filtered out the dipping intervals, these features were no longer present in the persistent spectra. 

\section*{Acknowledgements}
This study is based on the observations obtained with \emph{Suzaku}, a JAXA/ISAS space mission, and used the archived data and software provided by the High Energy Astrophysics Science Archive Research Center (HEASARC) online service maintained by the NASA Goddard Space Flight Center. PS is supported financially by the Council of Scientific \& Industrial Research (CSIR) under the Junior Research Fellowship (JRF) scheme. We thank the anonymous referee for useful suggestions and comments. PS acknowledges support from the \emph{Suzaku} help team for the helpful discussion on instrumentation.




\bibliographystyle{mnras}
\bibliography{example} 

\begin{thebibliography}{}
\makeatletter
\relax
\def\mn@urlcharsother{\let\do\@makeother \do\$\do\&\do\#\do\^\do\_\do\%\do\~}
\def\mn@doi{\begingroup\mn@urlcharsother \@ifnextchar [ {\mn@doi@}
  {\mn@doi@[]}}
\def\mn@doi@[#1]#2{\def\@tempa{#1}\ifx\@tempa\@empty \href
  {http://dx.doi.org/#2} {doi:#2}\else \href {http://dx.doi.org/#2} {#1}\fi
  \endgroup}
\def\mn@eprint#1#2{\mn@eprint@#1:#2::\@nil}
\def\mn@eprint@arXiv#1{\href {http://arxiv.org/abs/#1} {{\tt arXiv:#1}}}
\def\mn@eprint@dblp#1{\href {http://dblp.uni-trier.de/rec/bibtex/#1.xml}
  {dblp:#1}}
\def\mn@eprint@#1:#2:#3:#4\@nil{\def\@tempa {#1}\def\@tempb {#2}\def\@tempc
  {#3}\ifx \@tempc \@empty \let \@tempc \@tempb \let \@tempb \@tempa \fi \ifx
  \@tempb \@empty \def\@tempb {arXiv}\fi \@ifundefined
  {mn@eprint@\@tempb}{\@tempb:\@tempc}{\expandafter \expandafter \csname
  mn@eprint@\@tempb\endcsname \expandafter{\@tempc}}}

\bibitem[\protect\citeauthoryear{{Armas Padilla}, {Ueda}, {Hori}, {Shidatsu}
  \& {Mu{\~n}oz-Darias}}{{Armas Padilla} et~al.}{2017}]{Armas2017}
{Armas Padilla} M.,  {Ueda} Y.,  {Hori} T.,  {Shidatsu} M.,
  {Mu{\~n}oz-Darias} T.,  2017, \mn@doi [\mnras] {10.1093/mnras/stx020}, \href
  {https://ui.adsabs.harvard.edu/abs/2017MNRAS.467..290A} {467, 290}

\bibitem[\protect\citeauthoryear{{Arnaud}}{{Arnaud}}{1996}]{Arnaud1996}
{Arnaud} K.~A.,  1996, {XSPEC: The First Ten Years}.
p.~17

\bibitem[\protect\citeauthoryear{{Barret}, {Olive}, {Boirin}, {Done}, {Skinner}
   \& {Grindlay}}{{Barret} et~al.}{2000}]{Barret2000}
{Barret} D.,  {Olive} J.~F.,  {Boirin} L.,  {Done} C.,  {Skinner} G.~K.,
  {Grindlay} J.~E.,  2000, \mn@doi [\apj] {10.1086/308651}, \href
  {https://ui.adsabs.harvard.edu/abs/2000ApJ...533..329B} {533, 329}

\bibitem[\protect\citeauthoryear{{Barret}, {Olive}  \& {Oosterbroek}}{{Barret}
  et~al.}{2003}]{Barret2003}
{Barret} D.,  {Olive} J.~F.,   {Oosterbroek} T.,  2003, \mn@doi [\aap]
  {10.1051/0004-6361:20030011}, \href
  {https://ui.adsabs.harvard.edu/abs/2003A&A...400..643B} {400, 643}

\bibitem[\protect\citeauthoryear{{Boirin}, {M{\'e}ndez}, {D{\'\i}az Trigo},
  {Parmar}  \& {Kaastra}}{{Boirin} et~al.}{2005}]{Boirin2005}
{Boirin} L.,  {M{\'e}ndez} M.,  {D{\'\i}az Trigo} M.,  {Parmar} A.~N.,
  {Kaastra} J.~S.,  2005, \mn@doi [\aap] {10.1051/0004-6361:20041940}, \href
  {https://ui.adsabs.harvard.edu/abs/2005A&A...436..195B} {436, 195}

\bibitem[\protect\citeauthoryear{{Boldt}}{{Boldt}}{1987}]{Boldt1987}
{Boldt} E.,  1987, in {Hewitt} A.,  {Burbidge} G.,   {Fang} L.~Z.,  eds,  IAU
  Symposium Vol. 124, Observational Cosmology. p.~611

\bibitem[\protect\citeauthoryear{{Cackett}, {Altamirano}, {Patruno}, {Miller},
  {Reynolds}, {Linares}  \& {Wijnands}}{{Cackett} et~al.}{2009}]{Cackett2009}
{Cackett} E.~M.,  {Altamirano} D.,  {Patruno} A.,  {Miller} J.~M.,  {Reynolds}
  M.,  {Linares} M.,   {Wijnands} R.,  2009, \mn@doi [\apjl]
  {10.1088/0004-637X/694/1/L21}, \href
  {https://ui.adsabs.harvard.edu/abs/2009ApJ...694L..21C} {694, L21}

\bibitem[\protect\citeauthoryear{{Cackett} et~al.,}{{Cackett}
  et~al.}{2010}]{Cackett2010}
{Cackett} E.~M.,  et~al., 2010, \mn@doi [\apj] {10.1088/0004-637X/720/1/205},
  \href {https://ui.adsabs.harvard.edu/abs/2010ApJ...720..205C} {720, 205}

\bibitem[\protect\citeauthoryear{{Church} \& {Baluci{\'n}ska-Church}}{{Church}
  \& {Baluci{\'n}ska-Church}}{2001}]{Church2001}
{Church} M.~J.,  {Baluci{\'n}ska-Church} M.,  2001, \mn@doi [\aap]
  {10.1051/0004-6361:20010150}, \href
  {https://ui.adsabs.harvard.edu/abs/2001A&A...369..915C} {369, 915}

\bibitem[\protect\citeauthoryear{{Cottam}, {Kahn}, {Brinkman}, {den Herder}  \&
  {Erd}}{{Cottam} et~al.}{2001}]{Cottam2001}
{Cottam} J.,  {Kahn} S.~M.,  {Brinkman} A.~C.,  {den Herder} J.~W.,   {Erd} C.,
   2001, \mn@doi [\aap] {10.1051/0004-6361:20000053}, \href
  {https://ui.adsabs.harvard.edu/abs/2001A&A...365L.277C} {365, L277}

\bibitem[\protect\citeauthoryear{{Degenaar}, {Miller}, {Harrison}, {Kennea},
  {Kouveliotou}  \& {Younes}}{{Degenaar} et~al.}{2014}]{Degenaar2014}
{Degenaar} N.,  {Miller} J.~M.,  {Harrison} F.~A.,  {Kennea} J.~A.,
  {Kouveliotou} C.,   {Younes} G.,  2014, \mn@doi [\apjl]
  {10.1088/2041-8205/796/1/L9}, \href
  {https://ui.adsabs.harvard.edu/abs/2014ApJ...796L...9D} {796, L9}

\bibitem[\protect\citeauthoryear{{Di Salvo} et~al.,}{{Di Salvo}
  et~al.}{2015}]{Disalvo2015}
{Di Salvo} T.,  et~al., 2015, \mn@doi [\mnras] {10.1093/mnras/stv443}, \href
  {https://ui.adsabs.harvard.edu/abs/2015MNRAS.449.2794D} {449, 2794}

\bibitem[\protect\citeauthoryear{{D{\'\i}az Trigo} \& {Boirin}}{{D{\'\i}az
  Trigo} \& {Boirin}}{2016}]{Diaz2016}
{D{\'\i}az Trigo} M.,  {Boirin} L.,  2016, \mn@doi [Astronomische Nachrichten]
  {10.1002/asna.201612315}, \href
  {https://ui.adsabs.harvard.edu/abs/2016AN....337..368D} {337, 368}

\bibitem[\protect\citeauthoryear{{D{\'\i}az Trigo}, {Parmar}, {Boirin},
  {M{\'e}ndez}  \& {Kaastra}}{{D{\'\i}az Trigo} et~al.}{2006}]{Diaz2006}
{D{\'\i}az Trigo} M.,  {Parmar} A.~N.,  {Boirin} L.,  {M{\'e}ndez} M.,
  {Kaastra} J.~S.,  2006, \mn@doi [\aap] {10.1051/0004-6361:20053586}, \href
  {https://ui.adsabs.harvard.edu/abs/2006A&A...445..179D} {445, 179}

\bibitem[\protect\citeauthoryear{{Ebisawa}, {Bourban}, {Bodaghee}, {Mowlavi}
  \& {Courvoisier}}{{Ebisawa} et~al.}{2003}]{Ebisawa2003}
{Ebisawa} K.,  {Bourban} G.,  {Bodaghee} A.,  {Mowlavi} N.,   {Courvoisier}
  T.~J.~L.,  2003, \mn@doi [\aap] {10.1051/0004-6361:20031336}, \href
  {https://ui.adsabs.harvard.edu/abs/2003A&A...411L..59E} {411, L59}

\bibitem[\protect\citeauthoryear{{Gambino} et~al.,}{{Gambino}
  et~al.}{2019}]{Gambino2019}
{Gambino} A.~F.,  et~al., 2019, \mn@doi [\aap] {10.1051/0004-6361/201832984},
  \href {https://ui.adsabs.harvard.edu/abs/2019A&A...625A..92G} {625, A92}

\bibitem[\protect\citeauthoryear{{Gierli{\'n}ski} \&
  {Poutanen}}{{Gierli{\'n}ski} \& {Poutanen}}{2005}]{Gierlinski2005}
{Gierli{\'n}ski} M.,  {Poutanen} J.,  2005, \mn@doi [\mnras]
  {10.1111/j.1365-2966.2005.09004.x}, \href
  {https://ui.adsabs.harvard.edu/abs/2005MNRAS.359.1261G} {359, 1261}

\bibitem[\protect\citeauthoryear{{Iaria}, {Di Salvo}, {Burderi}  \&
  {Robba}}{{Iaria} et~al.}{2001}]{Iaria2001}
{Iaria} R.,  {Di Salvo} T.,  {Burderi} L.,   {Robba} N.~R.,  2001, \mn@doi
  [\apj] {10.1086/319010}, \href
  {https://ui.adsabs.harvard.edu/abs/2001ApJ...548..883I} {548, 883}

\bibitem[\protect\citeauthoryear{{Iaria} et~al.,}{{Iaria}
  et~al.}{2019}]{Iaria2019}
{Iaria} R.,  et~al., 2019, \mn@doi [\aap] {10.1051/0004-6361/201833982}, \href
  {https://ui.adsabs.harvard.edu/abs/2019A&A...630A.138I} {630, A138}

\bibitem[\protect\citeauthoryear{{Ibragimov} \& {Poutanen}}{{Ibragimov} \&
  {Poutanen}}{2009}]{Ibragimov2009}
{Ibragimov} A.,  {Poutanen} J.,  2009, \mn@doi [\mnras]
  {10.1111/j.1365-2966.2009.15477.x}, \href
  {https://ui.adsabs.harvard.edu/abs/2009MNRAS.400..492I} {400, 492}

\bibitem[\protect\citeauthoryear{{Jain} \& {Paul}}{{Jain} \&
  {Paul}}{2011}]{Jain2011}
{Jain} C.,  {Paul} B.,  2011, \mn@doi [\mnras]
  {10.1111/j.1365-2966.2010.18110.x}, \href
  {https://ui.adsabs.harvard.edu/abs/2011MNRAS.413....2J} {413, 2}

\bibitem[\protect\citeauthoryear{{Klu{\'z}niak} \& {Rappaport}}{{Klu{\'z}niak}
  \& {Rappaport}}{2007}]{Kluz2007}
{Klu{\'z}niak} W.,  {Rappaport} S.,  2007, \mn@doi [\apj] {10.1086/522954},
  \href {https://ui.adsabs.harvard.edu/abs/2007ApJ...671.1990K} {671, 1990}

\bibitem[\protect\citeauthoryear{{Kokubun} et~al.,}{{Kokubun}
  et~al.}{2007}]{Kokubun2007}
{Kokubun} M.,  et~al., 2007, \mn@doi [\pasj] {10.1093/pasj/59.sp1.S53}, \href
  {https://ui.adsabs.harvard.edu/abs/2007PASJ...59S..53K} {59, 53}

\bibitem[\protect\citeauthoryear{{Koyama} et~al.,}{{Koyama}
  et~al.}{2007}]{Koyama2007}
{Koyama} K.,  et~al., 2007, \mn@doi [\pasj] {10.1093/pasj/59.sp1.S23}, \href
  {https://ui.adsabs.harvard.edu/abs/2007PASJ...59S..23K} {59, 23}

\bibitem[\protect\citeauthoryear{{Lewin} \& {Clark}}{{Lewin} \&
  {Clark}}{1980}]{Lewin1980}
{Lewin} W.~H.~G.,  {Clark} G.~W.,  1980, in Ninth Texas Symposium on
  Relativistic Astrophysics. pp 451--478,
  \mn@doi{10.1111/j.1749-6632.1980.tb15953.x}

\bibitem[\protect\citeauthoryear{{Lewin} \& {van der Klis}}{{Lewin} \& {van der
  Klis}}{2006}]{vander2006}
{Lewin} W. H.~G.,  {van der Klis} M.,  2006, {Compact Stellar X-ray Sources}.
 Vol. 39

\bibitem[\protect\citeauthoryear{{Lin}, {Remillard}  \& {Homan}}{{Lin}
  et~al.}{2007}]{Lin2007}
{Lin} D.,  {Remillard} R.~A.,   {Homan} J.,  2007, \mn@doi [\apj]
  {10.1086/521181}, \href
  {https://ui.adsabs.harvard.edu/abs/2007ApJ...667.1073L} {667, 1073}

\bibitem[\protect\citeauthoryear{{Long}, {Romanova}  \& {Lovelace}}{{Long}
  et~al.}{2005}]{Long2005}
{Long} M.,  {Romanova} M.~M.,   {Lovelace} R.~V.~E.,  2005, \mn@doi [\apj]
  {10.1086/497000}, \href
  {https://ui.adsabs.harvard.edu/abs/2005ApJ...634.1214L} {634, 1214}

\bibitem[\protect\citeauthoryear{{Ludlam} et~al.,}{{Ludlam}
  et~al.}{2016}]{Ludlam2016}
{Ludlam} R.~M.,  et~al., 2016, \mn@doi [\apj] {10.3847/0004-637X/824/1/37},
  \href {https://ui.adsabs.harvard.edu/abs/2016ApJ...824...37L} {824, 37}

\bibitem[\protect\citeauthoryear{{Markwardt}, {Marshall}, {Swank}  \&
  {Takeshima}}{{Markwardt} et~al.}{1998}]{Markwardt1998}
{Markwardt} C.~B.,  {Marshall} F.~E.,  {Swank} J.,   {Takeshima} T.,  1998,
  \iaucirc, \href {https://ui.adsabs.harvard.edu/abs/1998IAUC.6998....2M}
  {6998, 2}

\bibitem[\protect\citeauthoryear{{Markwardt}, {Swank}  \&
  {Strohmayer}}{{Markwardt} et~al.}{2001}]{Markwardt2001}
{Markwardt} C.~B.,  {Swank} J.~H.,   {Strohmayer} T.~E.,  2001, in American
  Astronomical Society Meeting Abstracts. p. 27.04

\bibitem[\protect\citeauthoryear{{Mitsuda} et~al.,}{{Mitsuda}
  et~al.}{1984}]{Mitsuda1984}
{Mitsuda} K.,  et~al., 1984, \pasj, \href
  {https://ui.adsabs.harvard.edu/abs/1984PASJ...36..741M} {36, 741}

\bibitem[\protect\citeauthoryear{{Mitsuda}, {Inoue}, {Nakamura}  \&
  {Tanaka}}{{Mitsuda} et~al.}{1989}]{Mitsuda1989}
{Mitsuda} K.,  {Inoue} H.,  {Nakamura} N.,   {Tanaka} Y.,  1989, \pasj, \href
  {https://ui.adsabs.harvard.edu/abs/1989PASJ...41...97M} {41, 97}

\bibitem[\protect\citeauthoryear{{Mitsuda} et~al.,}{{Mitsuda}
  et~al.}{2007}]{Mitsuda2007}
{Mitsuda} K.,  et~al., 2007, \mn@doi [\pasj] {10.1093/pasj/59.sp1.S1}, \href
  {https://ui.adsabs.harvard.edu/abs/2007PASJ...59S...1M} {59, S1}

\bibitem[\protect\citeauthoryear{{Mukherjee}, {Bult}, {van der Klis}  \&
  {Bhattacharya}}{{Mukherjee} et~al.}{2015}]{Mukherjee2015}
{Mukherjee} D.,  {Bult} P.,  {van der Klis} M.,   {Bhattacharya} D.,  2015,
  \mn@doi [\mnras] {10.1093/mnras/stv1542}, \href
  {https://ui.adsabs.harvard.edu/abs/2015MNRAS.452.3994M} {452, 3994}

\bibitem[\protect\citeauthoryear{{Neilsen} \& {Lee}}{{Neilsen} \&
  {Lee}}{2009}]{Neilsen2009}
{Neilsen} J.,  {Lee} J.~C.,  2009, \mn@doi [\nat] {10.1038/nature07680}, \href
  {https://ui.adsabs.harvard.edu/abs/2009Natur.458..481N} {458, 481}

\bibitem[\protect\citeauthoryear{{Papitto}, {Riggio}, {di Salvo}, {Burderi},
  {D'A{\`\i}}, {Iaria}, {Bozzo}  \& {Menna}}{{Papitto}
  et~al.}{2010}]{Papitto2010}
{Papitto} A.,  {Riggio} A.,  {di Salvo} T.,  {Burderi} L.,  {D'A{\`\i}} A.,
  {Iaria} R.,  {Bozzo} E.,   {Menna} M.~T.,  2010, \mn@doi [\mnras]
  {10.1111/j.1365-2966.2010.17090.x}, \href
  {https://ui.adsabs.harvard.edu/abs/2010MNRAS.407.2575P} {407, 2575}

\bibitem[\protect\citeauthoryear{{Patruno}, {Wette}  \& {Messenger}}{{Patruno}
  et~al.}{2018}]{Patruno2018}
{Patruno} A.,  {Wette} K.,   {Messenger} C.,  2018, \mn@doi [\apj]
  {10.3847/1538-4357/aabf89}, \href
  {https://ui.adsabs.harvard.edu/abs/2018ApJ...859..112P} {859, 112}

\bibitem[\protect\citeauthoryear{{Pintore} et~al.,}{{Pintore}
  et~al.}{2016}]{Pintore2016}
{Pintore} F.,  et~al., 2016, \mn@doi [\mnras] {10.1093/mnras/stw176}, \href
  {https://ui.adsabs.harvard.edu/abs/2016MNRAS.457.2988P} {457, 2988}

\bibitem[\protect\citeauthoryear{{Ponti}, {Fender}, {Begelman}, {Dunn},
  {Neilsen}  \& {Coriat}}{{Ponti} et~al.}{2012}]{Ponti2012}
{Ponti} G.,  {Fender} R.~P.,  {Begelman} M.~C.,  {Dunn} R.~J.~H.,  {Neilsen}
  J.,   {Coriat} M.,  2012, \mn@doi [\mnras]
  {10.1111/j.1745-3933.2012.01224.x}, \href
  {https://ui.adsabs.harvard.edu/abs/2012MNRAS.422L..11P} {422, L11}

\bibitem[\protect\citeauthoryear{{Psaltis} \& {Chakrabarty}}{{Psaltis} \&
  {Chakrabarty}}{1999}]{Psaltis1999}
{Psaltis} D.,  {Chakrabarty} D.,  1999, \mn@doi [\apj] {10.1086/307525}, \href
  {https://ui.adsabs.harvard.edu/abs/1999ApJ...521..332P} {521, 332}

\bibitem[\protect\citeauthoryear{{Psaradaki}, {Costantini}, {Mehdipour}  \&
  {D{\'\i}az Trigo}}{{Psaradaki} et~al.}{2018}]{Psaradaki2018}
{Psaradaki} I.,  {Costantini} E.,  {Mehdipour} M.,   {D{\'\i}az Trigo} M.,
  2018, \mn@doi [\aap] {10.1051/0004-6361/201834000}, \href
  {https://ui.adsabs.harvard.edu/abs/2018A&A...620A.129P} {620, A129}

\bibitem[\protect\citeauthoryear{{Raman}, {Maitra}  \& {Paul}}{{Raman}
  et~al.}{2018}]{Raman2018}
{Raman} G.,  {Maitra} C.,   {Paul} B.,  2018, \mn@doi [\mnras]
  {10.1093/mnras/sty918}, \href
  {https://ui.adsabs.harvard.edu/abs/2018MNRAS.477.5358R} {477, 5358}

\bibitem[\protect\citeauthoryear{{Reeves}, {Done}, {Pounds}, {Terashima},
  {Hayashida}, {Anabuki}, {Uchino}  \& {Turner}}{{Reeves}
  et~al.}{2008}]{Reeves2008}
{Reeves} J.,  {Done} C.,  {Pounds} K.,  {Terashima} Y.,  {Hayashida} K.,
  {Anabuki} N.,  {Uchino} M.,   {Turner} M.,  2008, \mn@doi [\mnras]
  {10.1111/j.1745-3933.2008.00443.x}, \href
  {https://ui.adsabs.harvard.edu/abs/2008MNRAS.385L.108R} {385, L108}

\bibitem[\protect\citeauthoryear{{Sanna} et~al.,}{{Sanna}
  et~al.}{2018}]{Sanna2018}
{Sanna} A.,  et~al., 2018, \mn@doi [\aap] {10.1051/0004-6361/201732262}, \href
  {https://ui.adsabs.harvard.edu/abs/2018A&A...610L...2S} {610, L2}

\bibitem[\protect\citeauthoryear{{Shakura} \& {Sunyaev}}{{Shakura} \&
  {Sunyaev}}{1973}]{Shakura1973}
{Shakura} N.~I.,  {Sunyaev} R.~A.,  1973, \aap, \href
  {https://ui.adsabs.harvard.edu/abs/1973A&A....24..337S} {500, 33}

\bibitem[\protect\citeauthoryear{{Sharma}, {Jaleel}, {Jain}, {Pand ey}, {Paul}
  \& {Dutta}}{{Sharma} et~al.}{2018}]{Sharma2018}
{Sharma} R.,  {Jaleel} A.,  {Jain} C.,  {Pand ey} J.~C.,  {Paul} B.,   {Dutta}
  A.,  2018, \mn@doi [\mnras] {10.1093/mnras/sty2678}, \href
  {https://ui.adsabs.harvard.edu/abs/2018MNRAS.481.5560S} {481, 5560}

\bibitem[\protect\citeauthoryear{{Sharma}, {Jain}  \& {Dutta}}{{Sharma}
  et~al.}{2019}]{Sharma2019}
{Sharma} R.,  {Jain} C.,   {Dutta} A.,  2019, \mn@doi [\mnras]
  {10.1093/mnras/sty2808}, \href
  {https://ui.adsabs.harvard.edu/abs/2019MNRAS.482.1634S} {482, 1634}

\bibitem[\protect\citeauthoryear{{Sharma}, {Beri}, {Sanna}  \&
  {Dutta}}{{Sharma} et~al.}{2020}]{Sharma2020}
{Sharma} R.,  {Beri} A.,  {Sanna} A.,   {Dutta} A.,  2020, \mn@doi [\mnras]
  {10.1093/mnras/staa109}, \href
  {https://ui.adsabs.harvard.edu/abs/2020MNRAS.492.4361S} {492, 4361}

\bibitem[\protect\citeauthoryear{{Takahashi} et~al.,}{{Takahashi}
  et~al.}{2007}]{Takahashi2007}
{Takahashi} T.,  et~al., 2007, \mn@doi [\pasj] {10.1093/pasj/59.sp1.S35}, \href
  {https://ui.adsabs.harvard.edu/abs/2007PASJ...59S..35T} {59, 35}

\bibitem[\protect\citeauthoryear{{Verner}, {Ferland}, {Korista}  \&
  {Yakovlev}}{{Verner} et~al.}{1996}]{Verner1996}
{Verner} D.~A.,  {Ferland} G.~J.,  {Korista} K.~T.,   {Yakovlev} D.~G.,  1996,
  \mn@doi [\apj] {10.1086/177435}, \href
  {https://ui.adsabs.harvard.edu/abs/1996ApJ...465..487V} {465, 487}

\bibitem[\protect\citeauthoryear{{White} \& {Mason}}{{White} \&
  {Mason}}{1985}]{White1985}
{White} N.~E.,  {Mason} K.~O.,  1985, \mn@doi [\ssr] {10.1007/BF00212883},
  \href {https://ui.adsabs.harvard.edu/abs/1985SSRv...40..167W} {40, 167}

\bibitem[\protect\citeauthoryear{{Wilms}, {Allen}  \& {McCray}}{{Wilms}
  et~al.}{2000}]{Wilms2000}
{Wilms} J.,  {Allen} A.,   {McCray} R.,  2000, \mn@doi [\apj] {10.1086/317016},
  \href {https://ui.adsabs.harvard.edu/abs/2000ApJ...542..914W} {542, 914}

\bibitem[\protect\citeauthoryear{{Yoshida}, {Mitsuda}, {Ebisawa}, {Ueda},
  {Fujimoto}, {Yaqoob}  \& {Done}}{{Yoshida} et~al.}{1993}]{Yoshida1993}
{Yoshida} K.,  {Mitsuda} K.,  {Ebisawa} K.,  {Ueda} Y.,  {Fujimoto} R.,
  {Yaqoob} T.,   {Done} C.,  1993, \pasj, \href
  {https://ui.adsabs.harvard.edu/abs/1993PASJ...45..605Y} {45, 605}

\bibitem[\protect\citeauthoryear{{Younes}, {Boirin}  \& {Sabra}}{{Younes}
  et~al.}{2009}]{Younes2009}
{Younes} G.,  {Boirin} L.,   {Sabra} B.,  2009, \mn@doi [\aap]
  {10.1051/0004-6361/200811314}, \href
  {https://ui.adsabs.harvard.edu/abs/2009A&A...502..905Y} {502, 905}

\bibitem[\protect\citeauthoryear{{Zdziarski}, {Johnson}  \&
  {Magdziarz}}{{Zdziarski} et~al.}{1996}]{Zdziarski1996}
{Zdziarski} A.~A.,  {Johnson} W.~N.,   {Magdziarz} P.,  1996, \mn@doi [\mnras]
  {10.1093/mnras/283.1.193}, \href
  {https://ui.adsabs.harvard.edu/abs/1996MNRAS.283..193Z} {283, 193}

\bibitem[\protect\citeauthoryear{{Zhang}, {Makishima}, {Sakurai}, {Sasano}  \&
  {Ono}}{{Zhang} et~al.}{2014}]{Zhang2014}
{Zhang} Z.,  {Makishima} K.,  {Sakurai} S.,  {Sasano} M.,   {Ono} K.,  2014,
  \mn@doi [\pasj] {10.1093/pasj/psu117}, \href
  {https://ui.adsabs.harvard.edu/abs/2014PASJ...66..120Z} {66, 120}

\bibitem[\protect\citeauthoryear{{Zhang}, {Sakurai}, {Makishima}, {Nakazawa},
  {Ono}, {Yamada}  \& {Xu}}{{Zhang} et~al.}{2016}]{Zhang2016}
{Zhang} Z.,  {Sakurai} S.,  {Makishima} K.,  {Nakazawa} K.,  {Ono} K.,
  {Yamada} S.,   {Xu} H.,  2016, \mn@doi [\apj] {10.3847/0004-637X/823/2/131},
  \href {https://ui.adsabs.harvard.edu/abs/2016ApJ...823..131Z} {823, 131}

\bibitem[\protect\citeauthoryear{{{\.Z}ycki}, {Done}  \& {Smith}}{{{\.Z}ycki}
  et~al.}{1999}]{Zycki1999}
{{\.Z}ycki} P.~T.,  {Done} C.,   {Smith} D.~A.,  1999, \mn@doi [\mnras]
  {10.1046/j.1365-8711.1999.02885.x}, \href
  {https://ui.adsabs.harvard.edu/abs/1999MNRAS.309..561Z} {309, 561}

\bibitem[\protect\citeauthoryear{{in 't Zand} et~al.,}{{in 't Zand}
  et~al.}{1999}]{intzand1999}
{in 't Zand} J.~J.~M.,  et~al., 1999, \aap, \href
  {https://ui.adsabs.harvard.edu/abs/1999A&A...345..100I} {345, 100}

\bibitem[\protect\citeauthoryear{{van Peet}, {Costantini}, {M{\'e}ndez},
  {Paerels}  \& {Cottam}}{{van Peet} et~al.}{2009}]{Peet2009}
{van Peet} J.~C.~A.,  {Costantini} E.,  {M{\'e}ndez} M.,  {Paerels} F.~B.~S.,
  {Cottam} J.,  2009, \mn@doi [\aap] {10.1051/0004-6361/200811181}, \href
  {https://ui.adsabs.harvard.edu/abs/2009A&A...497..805V} {497, 805}

\makeatother
\end{thebibliography}







\bsp	
\label{lastpage}
\end{document}